\documentclass[a4paper, 10pt]{article}

\usepackage[margin=3.2cm]{geometry}

\usepackage{amsmath}
\usepackage{color}
\usepackage{bbm}

\newcommand{\bela}[1]{\begin{equation}\label{#1}}
\newcommand{\ela}{\end{equation}}
\newcommand{\vphi}{\varphi}
\newcommand{\bx}{\mbox{\boldmath$x$}}
\newcommand{\by}{\mbox{\boldmath$y$}}
\newcommand{\bz}{\mbox{\boldmath$z$}}
\newcommand{\bb}{\mbox{\boldmath$b$}}
\newcommand{\bbf}{\mbox{\boldmath$f$}}
\newcommand{\bc}{\mbox{\boldmath$c$}}
\newcommand{\bu}{\mbox{\boldmath$u$}}
\newcommand{\NU}{\mbox{\boldmath$0$}}
\newcommand{\del}{\partial}
\newcommand{\ci}{\mathrm{i}}
\newcommand{\eins}{\mathbbm{1}}
\newcommand{\high}[1]{{\color{black} #1}}

\numberwithin{equation}{section}

\begin{document}

%%%% Article title to be placed here
\title{Affine manifolds: The differential geometry of the multi-dimensionally consistent TED equation}

\author{%%%% Author details
W. K. Schief$^{1}$, U. Hertrich-Jeromin$^{2}$ and B. G. Konopelchenko$^{3}$
\bigskip\\
$^{1}$School of Mathematics and Statistics,\\
The University of New South Wales,
Sydney, NSW 2052, Australia\\[2mm]
$^{2}$Technische Universit\"at Wien,\\
Wiener Hauptstra\ss e 8-10/104,
A-1040 Wien, Austria\\[2mm]
$^{3}$INFN, Sezione di Lecce,\\
via per Arnesano, Lecce, 73100, Italy
}

\maketitle

%%%% Abstract text to be placed here %%%%%%%%%%%%
\begin{abstract}
It is shown that a canonical geometric setting of the integrable TED equation is a K\"ahlerian tangent bundle of an affine manifold. The  remarkable multi-dimensional consistency of this 4+4-dimensional dispersionless partial differential equation arises naturally in this context. In a particular 4-dimensional reduction, the affine manifolds turn out to be self-dual Einstein spaces of neutral signature governed by Pleba\'nski's first heavenly equation. In another reduction, the affine manifolds are Hessian, governed by compatible general heavenly equations. The Legendre invariance of the latter gives rise to a (dual) Hessian structure. Foliations of  affine manifolds in terms of self-dual Einstein spaces are also shown to arise in connection with a natural 5-dimensional reduction.
\end{abstract}
%%%%%%%%%%%%%%%%%%%%%%%%%%%%%%

\section{Introduction}

The notion of multi-dimensional consistency \cite{NijhoffWalker2001, BobenkoSuris2002,AdlerBobenkoSuris2012} has proven to be central in both the algebraic and geometric theory of discrete integrable systems. However, multi-dimensional consistency of partial differential equations has been explored to a much lesser extent and, to date, multi-dimensional consistency in its strict form appears to be a very rare phenomenon in connection with partial differential equations. A geometrically relevant three-dimensional equation which admits this property is given by
\bela{I1}
  \vphi_{x^1x^2x^3} = \frac{1}{2}\left(\frac{\vphi_{x^1x^2}\vphi_{x^1x^3}}{\vphi_{x^1}} + \frac{\vphi_{x^1x^2}\vphi_{x^2x^3}}{\vphi_{x^2}} + \frac{\vphi_{x^1x^3}\vphi_{x^2x^3}}{\vphi_{x^3}}\right).
\ela
Indeed, it is readily verfied that if one considers a function $\vphi = \vphi(x^1,\ldots,x^n)$ then the system of $n\choose 3$ copies of \eqref{I1} obtained by replacing $x^1,x^2,x^3$ by any three distinct independent variables $x^i,x^k,x^l$ is in involution in the sense of Riquier-Janet (Cartan-K\"ahler) \cite{Schwarz1984,Seiler2010} theory. Accordingly, this system of equations is compatible and may be solved simultaneously without having to constrain the Cauchy data associated with any individual equation of the form \eqref{I1}. It turns out that \eqref{I1} is but another avatar of Darboux's equation governing `symmetric' conjugate nets in a three-dimensional Euclidean space \cite{SchiefRogersTsarev1995}. Moreover, it appears in \cite{LobbNijhoff2009} in connection with Lagrangian multiforms.

A four-dimensional differential equation which has been shown to be multi-dimensionally consistent \cite{Bogdanov2015} and has come to be known as the general heavenly equation \cite{DoubrovFerapontov2010} is given by 
\bela{I3}
  c_1\Lambda_{x^1x^2}\Lambda_{x^3x^4} + c_2\Lambda_{x^2x^3}\Lambda_{x^1x^4} + c_3\Lambda_{x^3x^1}\Lambda_{x^2x^4} = 0,\qquad c_1 + c_2 + c_3 = 0.
\ela
It was originally derived \cite{Schief1999,Schief1996} as the continuum limit of the superposition principle associated with the B\"acklund transformation for the integrable (discrete) Tzitz\'eica equation of affine (discrete) differential geometry \cite{BobenkoSchief1999}. Its geometric significance was also established in \cite{Schief1999,Schief1996}. Indeed, it was shown that the general heavenly equation governs self-dual Einstein spaces in the same manner as Pleba\'nski's important first heavenly equation \cite{Plebanski1975}
\bela{I4}
  \Omega_{u\bar{u}}\Omega_{v\bar{v}} - \Omega_{u\bar{v}}\Omega_{v\bar{u}} = \kappa.
\ela
In fact, it has been demonstrated that the general heavenly equation constitutes a unique eigenfunction equation associated with any Lax representation of the governing equations of self-dual Einstein spaces \cite{KonopelchenkoSchiefSzereszewski2021}. 

It is interesting to note that the four-dimensional general heavenly equation itself may be decomposed into the four compatible copies of the three-dimensional dispersionless Hirota equation
\bela{I4a}
  (\mu_i-\mu_k)\Lambda_{x^l}\Lambda_{x^ix^k} + (\mu_k-\mu_l)\Lambda_{x^i}\Lambda_{x^kx^l} + (\mu_l-\mu_i)\Lambda_{x^k}\Lambda_{x^lx^i}  = 0,
\ela
where $i,k,l\in\{1,2,3,4\}$ are distinct and the constants $c_i$ have been parametrised according to
\bela{I4b}
  c_1 = c_{1234},\quad c_2 = c_{2314},\quad c_3 = c_{3124},\quad c_{iklm} = (\mu_i-\mu_k)(\mu_l-\mu_m).  
\ela
Indeed, the general heavenly equation \eqref{I3} is seen to be an algebraic consequence of the system \eqref{I4a} 
and the multi-dimensional consistency of the dispersionless Hirota equation \cite{Krynski2018} has been shown to be inherited from that of the general heavenly equation \cite{KonopelchenkoSchief2019}. The geometric relevance of the (compatible) dispersionless Hirota equation(s) in terms of Veronese webs and hyper-CR Einstein-Weyl structures has also been established  (see \cite{Krynski2018} and references therein). Moreover, the general heavenly equation constitutes a four-dimensional symmetry reduction of the 4+4-dimensional differential equation \cite{KonopelchenkoSchief2019}
\bela{I5}
  \varepsilon^{iklm}\Theta_{x^iy^k}\Theta_{x^ly^m} = 0,
\ela
where $\varepsilon^{iklm}$ is the totally anti-symmetric Levi-Civita symbol. Remarkably, \eqref{I5}, termed TED equation, has been proven in \cite{KonopelchenkoSchief2019} to be likewise multi-dimensionally consistent. The latter work mainly focuses on the algebraic (integrability) properties of the TED equation. Here, by contrast, we propose and examine a natural geometric setting in which the TED equation together with its multi-dimensional consistency arise.

Part of the symmetry structure of the TED equation indicates that the tangent bundle of an affine manifold constitutes a natural geometric setting in which the TED equation may be examined. The tangent bundle of an affine manifold comes with a natural complex structure which, in turn, allows one to define a (pseudo-)K\"ahler structure on the tangent bundle so that the tangent bundle becomes a K\"ahler manifold \cite{Shima2007}. It turns out that the TED equation is descriptive of a particular class of such K\"ahler manifolds. More precisely, these K\"ahler manifolds are governed by compatible copies of the TED equation. Thus, algebraic multi-dimensional consistency of the TED equation is an important ingredient which guarantees the consistency of the geometric construction of the class of K\"ahler manifolds under consideration. Moreover, it is demonstrated how the Lax representation of the TED equation arises naturally in this geometric setting. In particular, the spectral parameter may be generated via an additional symmetry of the TED equation. It is noted that this kind of phenomenon has been frequently observed in the geometric theory of integrable systems. For instance, the spectral parameter of the Lax pair for pseudospherical surfaces has its origin in a Lie point invariance of the underlying Gauss equation, namely the sine-Gordon equation \cite{RogersSchief2002}.

The TED equation admits a variety of dimensional reductions which have both mathematical and physical relevance \cite{KonopelchenkoSchief2019}. Here, we focus on three reductions which may be intepreted in terms of the geometry of the affine base manifold. Firstly, for an affine manifold $M$ of dimension $n\geq 4$, a canonical symmetry reduction leading to $n\choose 4$ compatible general heavenly equations of the type \eqref{I3} leads to a Hessian structure on $M$. This turns out to be a specific example of the general connection between Hessian structures on an affine manifold $M$ and K\"ahler structures on the tangent bundel $TM$ \cite{Shima2007}. Moreover, it is well known that affine structures arise in pairs related by Legendre transformations. Remarkably, the general heavenly equation admits a Legendre-type invariance which maps the affine structure to its dual. Accordingly, the two potentials which generate the Hessian metric and its dual on the affine manifold are governed by the same system of differential equations, namely the compatible general heavenly equations. Secondly, in the case of four-dimensional affine manifolds $M$, another canonical four-dimensional reduction leads to K\"ahler metrics on $M$ with the associated K\"ahler potentials being constrained by a differential equation of the form \eqref{I4}. Thus, remarkably, the affine manifolds $M$ constitute self-dual Einstein spaces with the signature of the K\"ahler metrics being neutral since $\kappa<0$. Finally, foliations of affine manifolds in the case $n=5$, wherein the leaves constitute self-dual Einstein spaces, are naturally obtained in the third canonical reduction associated with Husain-Park-type equations \cite{DoubrovFerapontov2010} which govern the foliations. 

It should be stressed that the geometric theory developed here may readily be extended to incorporate the $2m$+$2m$-dimensional TED equation ($m\geq2$)
\bela{I6}
  \varepsilon^{i_1k_1\cdots i_mk_m}\Theta_{x^{i_1}y^{k_1}}\cdots\Theta_{x^{i_m}y^{k_m}} = 0
\ela
which is likewise multi-dimensionally consistent. The equivalent of the first symmetry reduction then leads to Hessian manifolds $M$ of dimension $n\geq 2m$ governed by analogues of the general heavenly equation \eqref{I3} of degree $m$ \cite{Bogdanov2015}. The second dimensional reduction also admits an analogue for $n=2m$ which leads to a generalisation of self-dual Einstein spaces. The geometry of the latter is currently being investigated.

\section{Affine manifolds}

In the following, it is assumed that $M$ denotes an $n$-dimensional real affine manifold, that is, a differentiable manifold equipped with a flat and torsion-free connection $D$ \cite{Shima2007}. Affine manifolds come with natural (local) coordinates $\bx = (x^1,\ldots, x^n)^T$ for which the Christoffel symbols associated with $D$ vanish. All such affine coordinate systems are related by affine transformations
\bela{E1}
  \bx \mapsto T\bx + \bb,
\ela
where $T$ is a non-singular constant matrix and the vector $\bb$ is likewise constant. Hence, in the sequel, all geometric objects are to be invariant under the above group of transformations.

Any tangent vector $y$ on $M$ is of the form
\bela{E2}
  y = y^i\del_{x^i},
\ela
wherein Einstein's convention of summation over repeated indices has been adopted. Hence, $(x^1,\ldots,x^n,y^1,\ldots,y^n)$ constitute (local) coordinates of the tangent bundle $TM$ regarded as a $2n$-dimensional manifold. It is then readily verified that the coordinates $\by = (y^1,\ldots,y^n)^T$ transform according to
\bela{E3}
  \by \mapsto T\by.
\ela
The latter implies \cite{Shima2007} that there exists a natural complex structure on $TM$ encoded in the complex coordinates
\bela{E4}
  z^i = x^i + \ci y^i,\quad \bar{z}^i =x^i-\ci y^i.
\ela
Indeed, the complex structure is well defined since the maps
\bela{E5}
  \bz \mapsto T\bz + \bb,\quad \bar{\bz} \mapsto T\bar{\bz} + \bb
\ela
are (anti-)holomorphic.

\section{K\"ahler geometry}

Due to the existence of the complex structure on $TM$, it is natural \cite{Shima2007} to postulate the existence of a K\"ahler structure \cite{Kaehler1932} on $TM$. It is recalled that a K\"ahler structure on an even-dimensional manifold $\tilde{M}$ (such as $TM$) equipped with a complex structure is a real closed and non-degenerate two-form $\omega$ on $\tilde{M}$ which is compatible with the complex structure. The latter implies that $\omega$ is (locally) of the form
\bela{E6}
  \omega = \ci g_{i\bar{k}}dz^i\wedge d\bar{z}^k,\quad  g_{k\bar{i}} = \bar{g}_{i\bar{k}}.
\ela
As is customary, the rows and columns of the Hermitian matrix $(g_{i\bar{k}})$ are labelled by $1,\ldots,n$ and $\bar{1},\ldots,\bar{n}$ respectively. Furthermore, the condition $d\omega=0$ leads to the partial differential equations
\bela{E7}
  \frac{\del g_{i\bar{k}}}{\del z^l} = \frac{\del g_{l\bar{k}}}{\del z^i},\quad 
  \frac{\del g_{i\bar{k}}}{\del \bar{z}^l} = \frac{\del g_{i\bar{l}}}{\del \bar{z}^k}.
\ela
Application of the Poincar\'e lemma then implies that, locally, there exists a real potential $\Theta$ such that \cite{Kaehler1932,KobayashiNomizu1969}
\bela{E8}
  g_{i\bar{k}} = \Theta_{z^i\bar{z}^k},
\ela
where the subscripts on the K\"ahler potential $\Theta$ denote derivatives. Hence, the K\"ahler form $\omega$ becomes
\bela{E9}
  \omega = \ci \Theta_{z^i\bar{z}^k}dz^i\wedge d\bar{z}^k.
\ela

It is noted that a K\"ahler form is invariant under (anti-)holomorphic maps $\bz \mapsto \bbf(\bz)$, $\bar{\bz} \mapsto \overline{\bbf(\bz)}$. The same applies to the associated real K\"ahler metric which is (locally) defined by
\bela{E10}
  ds^2 = g_{i\bar{k}}dz^id\bar{z}^k = \Theta_{z^i\bar{z}^k}dz^id\bar{z}^k.
\ela
In particular, if $\tilde{M}=TM$, the definition of $\omega$ and the corresponding K\"ahler metric does not depend on the choice of the affine coordinates $x^i$ on $M$ as may be deduced from \eqref{E5}.
Therefore, $TM$ constitutes a K\"ahler manifold. It is observed that, here, we do not insist on the K\"ahler metric being Riemannian so that the Hermitian matrix $(g_{i\bar{k}})$ is not required to be positive definite. It is noted that a K\"ahler metric is usually defined as $2g_{i\bar{k}}dz^id\bar{z}^k$ but, in the current context, the above definition is more convenient.

In terms of real coordinates (such as the affine coordinates in the case $\tilde{M}=TM$), a K\"ahler form $\omega$ may be written as
\bela{E11}
  2\omega = dx_i\wedge dy^i + dx^i\wedge dy_i,
\ela
where $x_i=\Theta_{x^i}$ and $y_i = \Theta_{y^i}$ so that
\bela{E12}
  d\Theta  = x_idx^i + y_idy^i.
\ela
In fact, without reference to the K\"ahler potential, the functions $x_i$ and $y_i$ may be defined by the integrability condition
\bela{E13}
  dx_i\wedge dx^i + dy_i\wedge dy^i = 0
\ela
of \eqref{E12} which guarantees the existence of the potential $\Theta$. Thus, the pair \eqref{E11}, \eqref{E13} locally captures K\"ahler geometry. In this connection, it is observed that the triple \eqref{E11}-\eqref{E13} is invariant under the Legendre transformation
\bela{E14}
  (x^i,y^i) \leftrightarrow (x_i,y_i),\quad \Theta \rightarrow x_ix^i + y_iy^i -\Theta.
\ela
The analogue of this transformation in Hessian geometry \cite{Shima2007} is well known and will be shown to be relevant in \S 5.

\section{The geometry of the TED equation}

\subsection{Particular K\"ahler geometries}

In order to proceed, we now introduce the notion of horizontal vector fields and forms on $TM$. Thus, any vector field $a$ on $TM$ is of the form
\bela{E15}
  a = a^k\del_{x^k} +\tilde{a}^k\del_{y^k},
\ela
where $a^k = a^k(x^i,y^i)$ and $\tilde{a}^k = \tilde{a}^k(x^i,y^i)$, and it is termed horizontal if $\tilde{a}^k=0$. In particular, horizontal vector fields at $y^i=0$ may be identified with vector fields on the submanifold $M$ of $TM$. If we consider the class of volume forms of the type $c_0dx^1\wedge\cdots\wedge dx^n\wedge dy^1\wedge\cdots\wedge dy^n$ for any constant $c_0$ (which is invariant under the group of allowable affine transformations \eqref{E1}, \eqref{E3}) then the associated divergence of a vector field on $TM$ is given by 
\bela{E15a}
  \operatorname{div} a = \del_{x^k}a^k +\del_{y^k}\tilde{a}^k.
\ela
In particular, the divergence of a horizontal vector field reduces to
\bela{E15b}
  \operatorname{div} a = \del_{x^k}a^k.
\ela

Similarly, an $m$-form $\sigma$ on $TM$ is termed horizontal if it does not contain any differentials $dy^i$, that is, if it is of the form
\bela{E16}
  \sigma = \sigma_{i_1\cdots i_m}dx^{i_1}\wedge\cdots\wedge dx^{i_m}.
\ela
Accordingly, we denote the horizontal part of any vector field $a$ and form $\sigma$ by $a^h$ and $\sigma^h$ respectively so that 
\bela{E17}
  \sigma^h = 0\quad\Leftrightarrow\quad \sigma(a_1^h,\ldots,a_m^h) = 0
\ela
for all sets of vector fields $a_1,\ldots,a_m$. It is emphasised that the notion of horizontal vector fields and forms is coordinate independent. Accordingly, if $n\geq 4$, particular K\"ahler geometries on $TM$ may be defined by the constraint
\bela{E18}
  (\omega\wedge\omega)^h = 0
\ela
or, equivalently,
\bela{E19}
  \omega^h\wedge\omega^h = 0.
\ela
Another way of formulating the above restriction is
\bela{E20}
  \omega\wedge\omega\wedge dy^1\wedge\cdots\wedge dy^n = 0.
\ela

\subsection{The TED equation}

The parametrisation \eqref{E11} of the K\"ahler form $\omega$ shows that the constraint \eqref{E18} becomes
\bela{E21}
  (dx^k\wedge dy_k\wedge dx^m\wedge dy_m)^h = 0
\ela
which, in terms of the K\"ahler potential $\Theta$, reads
\bela{E22}
  \Theta_{x^iy^k}\Theta_{x^ly^m} dx^i\wedge dx^k\wedge dx^l\wedge dx^m = 0.
\ela
In the case $n=4$, the latter constitutes a single equation for $\Theta$, namely
\bela{E23}
  \varepsilon^{iklm}\Theta_{x^iy^k}\Theta_{x^ly^m} = 0,
\ela
where $\varepsilon^{iklm}$ is the totally anti-symmetric Levi-Civita symbol defined by $\varepsilon^{1234}=1$. Explicitly, one obtains
\bela{E24}
 \begin{aligned}
   &(\Theta_{x^1y^2} - \Theta_{x^2y^1})(\Theta_{x^3y^4} - \Theta_{x^4y^3})\\
+ & (\Theta_{x^2y^3} - \Theta_{x^3y^2})(\Theta_{x^1y^4} - \Theta_{x^4y^1})\\
+ & (\Theta_{x^3y^1} - \Theta_{x^1y^3})(\Theta_{x^2y^4} - \Theta_{x^4y^2}) = 0,
 \end{aligned}
\ela
which is the 4+4-dimensional integrable TED equation \cite{KonopelchenkoSchief2019}. It is readily verified that the TED equation is indeed invariant under affine transformations of the type \eqref{E1}, \eqref{E3}.

If $n>4$ then one obtains the overdetermined system of equations
\bela{E25}
  \varepsilon^{iklm\kappa_{5}\cdots\kappa_n}\Theta_{x^iy^k}\Theta_{x^ly^m} = 0
\ela
which consists of $n\choose4$ TED equations
\bela{E26}
 \begin{aligned}
   &(\Theta_{x^iy^k} - \Theta_{x^ky^i})(\Theta_{x^ly^m} - \Theta_{x^my^l})\\
+ & (\Theta_{x^ky^l} - \Theta_{x^ly^k})(\Theta_{x^iy^m} - \Theta_{x^my^i})\\
+ & (\Theta_{x^ly^i} - \Theta_{x^iy^l})(\Theta_{x^ky^m} - \Theta_{x^my^k}) = 0
 \end{aligned}
\ela
with the indices $i,k,l,m\in\{1,\ldots,n\}$ being distinct. Remarkably, it has been shown that these equations are in involution (i.e., compatible with all integrability conditions being satisfied) in the sense of Riquier-Janet (Cartan-K\"ahler) theory \cite{Schwarz1984,Seiler2010} and may therefore be solved simultaneously. Moreover, by construction, even though each individual TED equation \eqref{E26} is not invariant under \eqref{E1}, \eqref{E3}, the collection of compatible TED equations is.

\subsection{Integrability}

The TED equation has been shown to be integrable in the sense of being multi-dimensionally consistent, admitting a Lax representation \cite{KonopelchenkoSchief2019} and being amenable to the $\bar{\del}$-dressing method \cite{BogdanovKonopelchenko2019}. Here, it is demonstrated that the Lax representation naturally arises in the current geometric context. Thus, for brevity, we confine ourselves to the case $n=4$ and consider horizontal vector fields of the form
\bela{E27}
  X^i = X^{im}\del_{x^m},\quad X^{im} = -X^{mi}.
\ela
The extension to the case $n>4$ is evident. The matrix $(X^{im})$ is skew-symmetric and, hence, its determinant is the square of the Pfaffian \cite{Hirota2003}
\bela{E27a}
  \operatorname{pf}(X^{im}) = X^{12}X^{34} + X^{23}X^{14} + X^{31}X^{24}.
\ela
If the above Pfaffian vanishes then the rank of the matrix $(X^{im})$ is 2 so that only two of the vector fields $X^i$ are linearly independent. It should be stressed that the condition \eqref{E27}$_2$ is not preserved by the affine transformations \eqref{E1}, \eqref{E3}. However, the latter merely produces linear combinations of vector fields for which this condition is satisfied. More precisely, if the affine transformations of the coordinates are supplemented by
\bela{E27aa}
  X^i \mapsto c{T^i}_k X^k,\quad T = ({T^i}_k),
\ela
where $c$ is any non-vanishing constant, then invariance is restored. Indeed, if we set
\bela{E27ab}
  \tilde{x}^i = {T^i}_k x^k + b^i,\quad \tilde{X}^i = c {T^i}_kX^k 
\ela
then
\bela{E27ac}
  \tilde{X}^i = c{T^i}_kX^{kl}{T^m}_l\del_{\tilde{x}^m} =: \tilde{X}^{im}\del_{\tilde{x}^m}
\ela
so that $\tilde{X}^{im} = -\tilde{X}^{mi}$ as required.

Now, by definition of the commutator of vector fields, we conclude that
\bela{E27b}
 \begin{aligned}
  {}[X^i,X^k] & = {(X^{il}X^{km}_{x^l} - X^{kl}X^{im}_{x^l})}\del_{x^m}\\
  & = [-{(X^{ik}X^{lm} + X^{kl}X^{im} + X^{li}X^{km})}_{x^l}\\
  &\phantom{= [}  + {(X^{ik}X^{lm})}_{x^l} - X^{il}_{x^l}X^{km} + X^{kl}_{x^l}X^{im}]\del_{x^m},
 \end{aligned}
\ela
which may be summarised as
\bela{E27c}
 \begin{aligned}
 {}[X^i,X^k] & = - \varepsilon^{iklm}{[\operatorname{pf}(X^{pq})]}_{x^l}\del_{x^m} + (\del_{x^l}X^{ik}) X^l\\
 &\phantom{= }- X^{ik}(\operatorname{div} X^m)\del_{x^m} -(\operatorname{div} X^i)X^k + (\operatorname{div} X^k)X^i.
 \end{aligned}
\ela
Since the vector fields $X^i$ are horizontal, their divergence is given by
\bela{E28}
  \operatorname{div} X^i = \del_{x^m}X^{im}.
\ela
Accordingly, if the divergence of the vector fields and the associated Pfaffian vanish, that is,
\bela{E29}
  \operatorname{div} X^i = 0,\quad \operatorname{pf}(X^{im}) = 0
\ela
then the vector fields $X^i$ define a two-dimensional integrable distribution. In this case, the linear system of partial differential equations
\bela{E31}
  X^i\psi = 0
\ela
for a function $\psi$ on $TM$ has rank 2 and is compatible.

It is evident that the choice
\bela{E27d}
  X^i = X^{im}\del_{x^m},\quad X^{im} = \varepsilon^{iklm}\Theta_{x^ky^l}
\ela
constitutes a solution of the vanishing divergence condition \eqref{E29}$_1$ and the vanishing Pfaffian condition \eqref{E29}$_2$ becomes the TED equation \eqref{E24}. Accordingly, the commutator relations \eqref{E27c} reduce to 
\bela{E30}
  [X^i,X^k] = \varepsilon^{iklm}\Theta_{x^lx^py^m}X^p
\ela
modulo the TED equation and if we consider an affine transformation of the type \eqref{E1}, \eqref{E3} then it is readily verified that
\bela{E30b}
  X^i \mapsto  (\det T)^{-1} {T^i}_k X^k
\ela
preserves the particular form of the vector fields \eqref{E27d}. Furthermore, the TED equation \eqref{E24} admits the additional symmetry
\bela{E32}
  \bx \rightarrow \alpha \bx + \beta \by,\quad \by \rightarrow \gamma \bx + \delta \by,
\ela
where the constants $\alpha,\ldots,\delta$ are only constrained by $\alpha\delta -\beta\gamma\neq 0$. This invariance essentially acts on the vector fields \eqref{E27d} according to
\bela{E33}
  X^i \rightarrow X^i_\mu = X^{im}(\del_{x^m} + \mu \del_{y^m})
\ela
with the constant $\mu$ depending on $\alpha,\ldots,\delta$. We therefore conclude that the $\mu$-dependent linear system
\bela{E34}
  X_{\mu}^i\psi = 0
\ela
is of rank 2 and compatible modulo the TED equation. It is emphasised that the vector fields $X^i_{\mu}$ are not horizontal unless $\mu=0$, in which case the vector fields $X^i$ are retrieved. However, these vector fields are divergence free for all $\mu$. In terms of integrable systems, \eqref{E34} constitutes a Lax pair \cite{ZakharovShabat1979} for the TED equation with $\mu$ representing the spectral parameter. 

We conclude this section by observing that, since the invariance \eqref{E32} is purely algebraic, the `complexification' $(x^i,y^i)\rightarrow(x^i+\ci y^i,x^i-\ci y^i)$ is also admissible. Accordingly, the TED equation admits the formulation
\bela{E36}
    \varepsilon^{iklm}\Theta_{z^i\bar{z}^k}\Theta_{z^l\bar{z}^m} = 0,
\ela
that is,
\bela{E37}
 \begin{aligned}
   &(\Theta_{z^1\bar{z}^2} - \Theta_{z^2\bar{z}^1})(\Theta_{z^3\bar{z}^4} - \Theta_{z^4\bar{z}^3})\\
+ & (\Theta_{z^2\bar{z}^3} - \Theta_{z^3\bar{z}^2})(\Theta_{z^1\bar{z}^4} - \Theta_{z^4\bar{z}^1})\\
+ & (\Theta_{z^3\bar{z}^1} - \Theta_{z^1\bar{z}^3})(\Theta_{z^2\bar{z}^4} - \Theta_{z^4\bar{z}^2}) = 0.
 \end{aligned}
\ela
Hence, in the case $n=4$, the particular K\"ahler manifolds considered here are directly encoded in the K\"ahler form \eqref{E9} (or K\"ahler metric \eqref{E10}), wherein the K\"ahler potential $\Theta$ is constrained by the above `complex' TED equation. For $n>4$, the K\"ahler potential obeys a compatible system of TED equations of the form \eqref{E37}.

\section{Hessian manifolds. The general heavenly equation}

The TED equation admits a variety of integrable reductions. Here, and in the following section, we focus on reductions which lead to interesting geometric properties of the affine manifold $M$. In arbitrary dimensions $n\geq 4$, a canonical symmetry reduction of the compatible system of TED equations \eqref{E26} is provided by assuming that $\Theta$ locally only depends on the $n$ components of a quantity of the form
\bela{E48a}
  \bx + S\by,
\ela
where $S$ is a constant matrix. The requirement that $S$ be diagonalisable with distinct real eigenvalues is also invariant under the group of affine transformations \eqref{E1}, \eqref{E3}. In this case, locally, there exist particular affine coordinates which are such that
\bela{E49}
   \Theta = \Lambda(s^i),\quad s^ i = x^i + \epsilon\mu_i y^i,
\ela
where $\epsilon\mu_i$ are the eigenvalues of $S$. The system \eqref{E26} (divided by $\epsilon^2$) then reduces to the system of compatible general heavenly equations
\bela{E50}
 \begin{aligned}
  (\mu_k-\mu_i)(\mu_m-\mu_l)\Lambda_{s^is^k}\Lambda_{s^ls^m} &\\
  + (\mu_l-\mu_k)(\mu_m-\mu_i)\Lambda_{s^ks^l}\Lambda_{s^is^m} &\\
  + (\mu_i-\mu_l)(\mu_m-\mu_k)\Lambda_{s^ls^i}\Lambda_{s^ks^m} & = 0,
 \end{aligned}
\ela
wherein the indices on the real constants $\mu_i$ are not subject to Einstein's summation convention. Since any function $\Theta$ which only depends on $\bx$ constitutes a solution of the TED equations \eqref{E26}, the parameter $\epsilon$ has been introduced to provide a meaningful limit as the eigenvalues $\epsilon\mu_i$ simultaneously tend to zero.  

The variables $s^i$ may be regarded as the real parts of the (anti-)holomorphic coordinates
\bela{E51}
  v^i = s^i + \ci t^i = (1-\ci\epsilon\mu_i)z^i,\quad \bar{v}^i = s^i - \ci t^i = (1+\ci\epsilon\mu_i)\bar{z}^i
\ela
and, hence, the associated K\"ahler metric \high{\eqref{E10}} is given by
\bela{E52}
  ds^2 = \tfrac{1}{4}\Lambda_{s^is^k}dv^id\bar{v}^k = \tfrac{1}{4}\Lambda_{s^is^k}(ds^ids^k + dt^idt^k).
\ela
On the submanifold $M$ on which $s^i=x^i$ and $t^i=-\epsilon\mu_ix^i$, the K\"ahler metric reduces to
\bela{E53}
  ds^2|_M = (1+\epsilon^2\mu_i\mu_k)\vphi_{x^ix^k}dx^idx^k,
\ela
where $\vphi=\frac{1}{4}\Lambda$ is governed by the compatible system of general heavenly equations \cite{Schief1999,Schief1996}
\bela{E54}
 \begin{aligned}
  (\mu_k-\mu_i)(\mu_m-\mu_l)\vphi_{x^ix^k}\vphi_{x^lx^m} &\\
  + (\mu_l-\mu_k)(\mu_m-\mu_i)\vphi_{x^kx^l}\vphi_{x^ix^m} &\\
  + (\mu_i-\mu_l)(\mu_m-\mu_k)\vphi_{x^lx^i}\vphi_{x^kx^m} & = 0.
 \end{aligned}
\ela
One may regard \eqref{E53} as a `deformed Hessian metric' on a `deformed Hessian manifold' $M$. Indeed, in the limit $\epsilon\rightarrow0$, we obtain
\bela{E54a}
 ds^2|_M\rightarrow \vphi_{x^ix^k}dx^idx^k.
\ela
Accordingly, in this case, $M$ constitutes a Hessian manifold \cite{Shima2007}. The latter is defined as an affine manifold equipped with a (pseudo-)Riemannian metric of the (local) form $D^2\vphi$, that is, \eqref{E54a} in terms of affine coordinates $x^i$. The metric \eqref{E54a} is indeed invariant under the group of affine transformations \eqref{E1}. In fact, the preceding may be regarded as a concrete illustration of a well-known general connection between a Hessian structure on $M$ and a K\"ahler structure on $TM$ \cite{Shima2007}. It should also be pointed out that particular (deformed) Hessian metrics are obtained by considering the decomposition of the general heavenly equations into compatible dispersionless Hirota equations as exemplified by the system \eqref{I4a} in the case $n=4$. It is observed that, in general, the K\"ahler metric \eqref{E52} on $TM$ may be reconstructed from the pair \eqref{E53}, \eqref{E54} determining the metric on $M$.

The general heavenly equation is known to admit a remarkable invariance of Legendre type \cite{KonopelchenkoSchiefSzereszewski2021}. Thus, if we introduce on $M$ the notation $\tilde{x}_i = \vphi_{x^i}$ as in \eqref{E12}, that is,
\bela{E55}
  d\vphi = \tilde{x}_idx^i
\ela
then, in analogy with \eqref{E21}, the system of general heavenly equations \eqref{E54} is encoded in the pair
\bela{E56}
  d\tilde{x}_i\wedge dx^i = 0,\quad \mu_k\mu_m dx^k\wedge d\tilde{x}_k\wedge dx^m\wedge d\tilde{x}_m = 0.
\ela
The latter is evidently symmetric in the variables $x^i$ and $\tilde{x}_i$ so that the Legendre transformation
\bela{E57}
  x^i\leftrightarrow \tilde{x}_i,\quad \vphi\rightarrow\tilde{\vphi}=\tilde{x}_ix^i - \vphi 
\ela
maps any solution $\vphi(x^i)$ of the system \eqref{E54} of general heavenly equations to another solution $\tilde{\vphi}(\tilde{x}_i)$. Accordingly, one may (locally) define another (pseudo-)Riemannian metric on $M$ by
\bela{E58}
  d\tilde{s}^2|_M = (1+\epsilon^2\mu_i\mu_k)\tilde{\vphi}_{\tilde{x}_i\tilde{x}_k}d\tilde{x}_i d\tilde{x}_k.
\ela
With regard to the limit $\epsilon\rightarrow0$, it is observed that, in fact, a general theorem in Hessian geometry \cite{Shima2007} states that, for any potential $\vphi$, the variables $\tilde{x}_i$ constitute affine coordinates associated with another (dual) flat and torsion-free connection $\tilde{D}=2\nabla - D$, where $\nabla$ is the Levi-Civita connection associated with the metric \eqref{E54a} and $D$ is the original affine connection. Hence, in the current situation, the two mutually dual Hessian structures are governed by the same system of (general heavenly) equations.

\section{Self-dual Einstein spaces. The Pleba\'nski equation}

Another canonical reduction of the TED equation \eqref{E24} which is invariant under \eqref{E1}, \eqref{E3} is given by
\bela{Z1}
  \Theta = \Omega(\bx + S\by) + (\bx+\bc)^TA\by,
\ela
where $A= - A^T$ is a constant skew-symmetric matrix and $S$ is now assumed to be diagonalisable with eigenvalues $\pm i$ of multiplicity 2. The vector $\bc$ is likewise constant. It is noted that the skew-symmetry of $A$ is motivated by the fact that any symmetric part of $A$ does not appear in the TED equation. Moreover, it is important to observe that $\Theta$ and $\Omega$ coincide on the manifold $M$, that is,
\bela{Z2}
  \Theta|_M = \Omega|_M.
\ela
The above assumptions may now be exploited by introducing adapted affine coordinates. Indeed, locally, there exist particular affine coordinates such that $S$ is of the form
\bela{Z3}
  S = \begin{pmatrix} \sigma&0\\ 0&\sigma\end{pmatrix}, \quad \sigma = \begin{pmatrix} 0 & -1\\ 1& 0\end{pmatrix}
\ela
and $\bc=\NU$. 

Based on the above adapted affine coordinates, it is now natural to introduce a new set of (anti-)holomorphic coordinates $u^i$ and $\bar{u}^i$ defined by
\bela{E35}
 \begin{aligned}
  &u^1 = z^1 + \ci z^2,\quad u^2 = z^1 - \ci z^2,\quad u^3 = z^3 + \ci z^4,\quad u^4 = z^3 - \ci z^4\\
  &\bar{u}^1 = \bar{z}^1 - \ci \bar{z}^2,\quad \bar{u}^2 = \bar{z}^1 + \ci \bar{z}^2,\quad
  \bar{u}^3 = \bar{z}^3 - \ci \bar{z}^4,\quad \bar{u}^4 = \bar{z}^3 + \ci \bar{z}^4.
\end{aligned}
\ela
In terms of these complex coordinates, the TED equation reads 
\bela{E38}
  (\Theta_{u^1\bar{u}^1}-\Theta_{u^2\bar{u}^2})(\Theta_{u^3\bar{u}^3}-\Theta_{u^4\bar{u}^4}) - |\Theta_{u^2\bar{u}^4}-\Theta_{u^3\bar{u}^1}|^2 + |\Theta_{u^3\bar{u}^2}-\Theta_{u^1\bar{u}^4}|^2 = 0.
\ela
By virtue of the algebraic nature of the invariance \eqref{E32}, it is seen that \eqref{E38} is conveniently obtained by making the formal substitution
\bela{E35a}
 \begin{aligned}
  &z^1\rightarrow u^1,\quad z^2 \rightarrow u^2,\quad z^3 \rightarrow u^3,\quad z^4 \rightarrow u^4\\
  &\bar{z}^1 \rightarrow \bar{u}^2,\quad \bar{z}^2 \rightarrow \bar{u}^1,\quad
  \bar{z}^3 \rightarrow \bar{u}^4,\quad \bar{z}^4 \rightarrow \bar{u}^3
\end{aligned}
\ela
in \eqref{E37}. Thus, we are left with the reduction
\bela{Z4}
  \Theta = \Omega(u^1,\bar{u}^1,u^3,\bar{u}^3) + \bx^TA\by
\ela
of the TED equation in the form \eqref{E38}. Here, the affine coordinates $x^i$ and $y^i$ need to be expressed in terms of the complex coordinates $u^i$ and $\bar{u}^i$. By construction, the associated K\"ahler metric is given by
\bela{E39}
  ds^2 = \Theta_{u^i\bar{u}^k}du^id\bar{u}^k.
\ela

It is natural to investigate whether the constant matrix $A$ may be constrained in any meaningful manner. To this end, it is emphasised that for any restriction to be `geometric', the constraint on $A$ must be invariant under the remaining subgroup of transformations \eqref{E1}, \eqref{E4} which preserves the normalised form \eqref{Z3} of $S$. This subgroup corresponds to matrices of the form
\bela{Z5}
  B = \begin{pmatrix} \sigma_{11}&\sigma_{12}\\ \sigma_{21}& \sigma_{22}\end{pmatrix},\quad [\sigma_{ik},\sigma] = 0.
\ela
The commutativity constraint on the matrices $\sigma_{ik}$ is resolved by the equivalent condition that $\sigma_{ik}\in\operatorname{span}\{\eins,\sigma\}$. One may now readily verify that the form of the block matrix
\bela{Z6}
  A = \begin{pmatrix} 0&Q\\ -Q&0\end{pmatrix},\quad Q^T = Q,\quad \operatorname{tr}Q = 0
\ela
is preserved by the transformation
\bela{EZ7}
  A\rightarrow B^TAB
\ela
so that it is geometrically admissible to impose the constraint \eqref{Z6} on the ansatz \eqref{Z4}.

In terms of the complex coordinates $u^i$ and $\bar{u}^i$, the parametrisation
\bela{Z8}
  Q = \begin{pmatrix}\hat{\nu} & \hat{\mu}\\ \hat{\mu}&-\hat{\nu}\end{pmatrix}
\ela
with $\tilde{\mu}$ and $\tilde{\nu}$ being constant now reveals that \eqref{Z4} or, equivalently,
\bela{Z9}
  \Theta = \Omega(u^1,\bar{u}^1,u^3,\bar{u}^3) + \frac{\ci}{2}\bz^TA\bar{\bz}
\ela
has the simple form
\bela{Z9a}
  \Theta = \Omega(u^1,\bar{u}^1,u^3,\bar{u}^3) + \bu^T\hat{A}\bar{\bu},\quad \hat{A}=\begin{pmatrix} 0&0&0&\alpha\\ 0&0&-\bar{\alpha}&0\\ 0&-\alpha&0&0\\ \bar{\alpha}&0&0&0\end{pmatrix},\quad \alpha = \frac{1}{4}(\hat{\mu}+\ci\hat{\nu}).
\ela
Explicit evaluation of the above leads to
\bela{E40}
  \Theta = \Omega(u^1,\bar{u}^1,u^3,\bar{u}^3) + \alpha (u^1\bar{u}^4  - u^3\bar{u}^2) + \bar{\alpha} (u^4\bar{u}^1 - u^2\bar{u}^3).
\ela

Insertion of the ansatz \eqref{E40} into the TED equation in the form \eqref{E38} now produces the four-dimensional partial differential equation
\bela{E41}
  \Omega_{u^1\bar{u}^1}\Omega_{u^3\bar{u}^3} - \Omega_{u^1\bar{u}^3}\Omega_{u^3\bar{u}^1} = -4|\alpha|^2
\ela
with associated K\"ahler metric
\bela{E42}
 \begin{aligned}
  ds^2 &= g_{i\bar{k}}du^id\bar{u}^k\\
&= \Omega_{u^1\bar{u}^1}du^1d\bar{u}^1 + \Omega_{u^3\bar{u}^3}du^3d\bar{u}^3 + \Omega_{u^1\bar{u}^3}du^1d\bar{u}^3 + \Omega_{u^3\bar{u}^1}du^3d\bar{u}^1\\
&\phantom{=}+ \alpha (du^1d\bar{u}^4  - du^3d\bar{u}^2) + \bar{\alpha} (du^4d\bar{u}^1 - du^2d\bar{u}^3).
 \end{aligned} 
\ela
In order to determine the signature of this metric, we note that, due to the block structure of any K\"ahler metric, its characteristic polynomial is given by
\bela{E43}
  [\det(\lambda\delta_{i\bar{k}} - g_{i\bar{k}})]^2,
\ela
where the Kronecker symbol $\delta_{i\bar{k}}$ is 1 for $i=k$ and 0 otherwise. Thus, in the current situation, there exist four eigenvalues $\lambda_i$ of multiplicity 2 and
\bela{E44}
  \lambda_1\lambda_2\lambda_3\lambda_4 = \det(g_{i\bar{k}}) = |\alpha|^4 = \mbox{const}.
\ela
Accordingly, the signature of the K\"ahler metric is (anti-)Riemannian $(\pm\pm\pm\pm\pm\pm\pm\pm)$ or neutral $(++++----)$. Moreover, it is known that the non-vanishing components of the Ricci tensor associated with a K\"ahler metric read \cite{Shima2007,Kaehler1932}
\bela{E45}
  R_{l\bar{m}} = {[\ln\det(g_{i\bar{k}})]}_{u^l\bar{u}^m}.
\ela
Therefore, the K\"ahler metric \eqref{E42} has vanishing Ricci tensor and Einstein's vacuum equations in 4+4 dimensions are satisfied.

The vanishing of the Ricci tensor is due to the algebraic structure of the K\"ahler metric \eqref{E42} and does not depend on  $\Omega$ being a solution of the differential equation \eqref{E41}. However, the relevance of the latter becomes apparent when one considers the restriction of the K\"ahler metric to the submanifold $M$ given by $y^i=0$ or, equivalently, $u^2=\bar{u}^1$ and $u^4=\bar{u}^3$. Indeed, since the metric on $M$ is
\bela{E46}
  ds^2|_M = \Omega_{u^1\bar{u}^1}du^1d\bar{u}^1 + \Omega_{u^3\bar{u}^3}du^3d\bar{u}^3 + \Omega_{u^1\bar{u}^3}du^1d\bar{u}^3 + \Omega_{u^3\bar{u}^1}du^3d\bar{u}^1,
\ela
$\det({g_{i\bar{k}}|}_M)$ coincides with the left-hand side of \eqref{E41} and is a negative constant. The Ricci tensor of the metric \eqref{E46} therefore vanishes and its signature is neutral. In fact, it is well known that pseudo-Riemannian manifolds equipped with a metric of the form \eqref{E46} and underlying differential equation \eqref{E41} represent self-dual Einstein spaces with \eqref{E41} being Pleba\'nski's first heavenly equation \cite{Plebanski1975}. 

In summary, it has been shown that the affine manifold $M$ becomes a self-dual Einstein space in the canonical reduction \eqref{E40} of the avatar \eqref{E38} of the TED equation. The metric \eqref{E42} on $TM$ may be reconstructed from \eqref{E46}. 

\section{Foliated manifolds}

In \S 5, it has been demonstrated how the multi-dimensional consistency of the TED equation guarantees the well-defined construction of (deformed) Hessian manifolds $M$ via the multi-dimensionally consistent general heavenly equation. It appears that multiple copies of Pleba\'nski's first heavenly equation \eqref{E41} are not compatible in a non-trivial sense. However, it has been shown in \cite{KonopelchenkoSchief2019} that the Pleban\'ski equation is compatible with four equations of Husain-Park type \cite{DoubrovFerapontov2010}. It turns out that this purely algebraic observation may be interpreted geometrically in the current context. This is achieved by considering a `hybrid' of the two canonical reductions analysed in the preceding. Thus, in the case $n=5$, we focus on reductions of the form \eqref{Z1}, where the matrix $S$ has eigenvalues $\pm\ci$ of multiplicity 2 and a real eigenvalue $\mu_5$. Accordingly, there exist particular affine coordinates such that S admits the normal form
\bela{F1}
  S = \begin{pmatrix} \sigma&0&0\\ 0&\sigma&0\\0&0&\mu_5\end{pmatrix}
\ela
and $\bc=\NU$. Furthermore, it is geometrically admissible to impose the form
\bela{F2}
  A = \begin{pmatrix} 0&Q&0\\ -Q&0&0\\ 0&0&0\end{pmatrix}
\ela
on the skew-symmetric matrix $A$. In the above, $\sigma$ and $Q$ are defined as in \eqref{Z3} and \eqref{Z6} respectively. In terms of the (anti-)holomorphic coordinates $u^i$ and $\bar{u}^i$ given by \eqref{E35} and
\bela{F3}
  s^5 = x^5 + \mu_5y^5,
\ela
the constraint on $\Theta$ becomes
\bela{F4}
  \Theta = \Omega(u^1,\bar{u}^1,u^3,\bar{u}^3,s^5) + \alpha (u^1\bar{u}^4  - u^3\bar{u}^2) + \bar{\alpha} (u^4\bar{u}^1 - u^2\bar{u}^3)
\ela
with the complex constant $\alpha$ being defined by \eqref{Z8}, \eqref{Z9a}$_3$.

The K\"ahler metric corresponding to the reduction \eqref{F4} is obtained best by, once again, regarding the variable $s^5$ as the real part of the (anti-)holomorphic coordinates
\bela{F5}
  u^5 = s^5 + \ci t^5 = (1-\ci\mu_5)z^5,\quad \bar{u}^5 = s^5 - \ci t^5 = (1+\ci\mu_5)\bar{z}^5
\ela
so that
\bela{F6}
 \begin{aligned}
  ds^2 &= g_{i\bar{k}}du^id\bar{u}^k\\
&= \Omega_{u^1\bar{u}^1}du^1d\bar{u}^1 + \Omega_{u^3\bar{u}^3}du^3d\bar{u}^3 + \Omega_{u^1\bar{u}^3}du^1d\bar{u}^3 + \Omega_{u^3\bar{u}^1}du^3d\bar{u}^1\\
&\phantom{=}+ \alpha (du^1d\bar{u}^4  - du^3d\bar{u}^2) + \bar{\alpha} (du^4d\bar{u}^1 - du^2d\bar{u}^3)\\
&\phantom{=}+\tfrac{1}{2}\Omega_{u^1s^5}du^1d\bar{u}^5 + \tfrac{1}{2}\Omega_{\bar{u}^1s^5}du^5d\bar{u}^1+\tfrac{1}{2}\Omega_{u^3s^5}du^3d\bar{u}^5 + \tfrac{1}{2}\Omega_{\bar{u}^3s^5}du^5d\bar{u}^3\\
&\phantom{-} + \tfrac{1}{4}\Omega_{s^5s^5}du^5d\bar{u}^5.
 \end{aligned} 
\ela
The function $\Omega$ obeys five differential equations generated by inserting the ansatz \eqref{F4} into five compatible copies of the TED equation in the form \eqref{E37}. By construction, the TED equation \eqref{E37} reduces to the Pleba\'nski equation \eqref{E41} and the other four equations are obtained by replacing in \eqref{E37} any of the indices $1,2,3,4$ by $5$ and applying the reduction \eqref{F4}. For brevity, we here merely state that these equations are given by
\bela{F7}
 \begin{aligned}
  \Omega_{u^1s^5} & = \bar{\rho}(\Omega_{u^1\bar{u}^3}\Omega_{\bar{u}^1s^5} - \Omega_{u^1\bar{u}^1}\Omega_{\bar{u}^3s^5})\\
  \Omega_{\bar{u}^1s^5} & = \rho(\Omega_{u^3\bar{u}^1}\Omega_{u^1s^5} - \Omega_{u^1\bar{u}^1}\Omega_{u^3s^5})\\
  \Omega_{u^3s^5} & = \bar{\rho}(\Omega_{u^3\bar{u}^3}\Omega_{\bar{u}^1s^5} - \Omega_{u^3\bar{u}^1}\Omega_{\bar{u}^3s^5})\\
  \Omega_{\bar{u}^3s^5} & = \rho(\Omega_{u^3\bar{u}^3}\Omega_{u^1s^5} - \Omega_{u^1\bar{u}^3}\Omega_{u^3s^5})
 \end{aligned}
\ela
with $2\alpha\rho= (\ci-\mu_5)/(\ci+\mu_5)$. These are the Husain-Park-type equations alluded to at the beginning of this section. It is noted that any three of the five equations \eqref{E41}, \eqref{F7} imply the
other two equations. This property is inherited from the fact that only three of the original five compatible TED equations are algebraically independent as recorded in \cite{KonopelchenkoSchief2019}. In particular, if one regards \eqref{F7} as a homogeneous linear system for the quantities $\Omega_{u^1s^5}$, $\Omega_{\bar{u}^1s^5}$, $\Omega_{u^3s^5}$ and $\Omega_{\bar{u}^3s^5}$ then the associated vanishing determinant is precisely the Pleba\'nski equation \eqref{E41}.

On the affine manifold $M$, corresponding to $u^2=\bar{u}^1$, $u^4=\bar{u}^3$ and $u^5 = (1-\ci\mu_5)s^5$, the metric \high{\eqref{F6}} reduces to
\bela{F8}
 \begin{aligned}
 ds^2|_M&= \Omega_{u^1\bar{u}^1}du^1d\bar{u}^1 + \Omega_{u^3\bar{u}^3}du^3d\bar{u}^3 + \Omega_{u^1\bar{u}^3}du^1d\bar{u}^3 + \Omega_{u^3\bar{u}^1}du^3d\bar{u}^1\\
&\phantom{=}+\tfrac{1}{2}[(1+\ci\mu_5)(\Omega_{u^1s^5}du^1 + \Omega_{u^3s^5}du^3) + (1-\ci\mu_5)(\Omega_{\bar{u}^1s^5}d\bar{u}^1+ \Omega_{\bar{u}^3s^5}d\bar{u}^3)]ds^5\\
&\phantom{-} + \tfrac{1}{4}(1+\mu_5^2)\Omega_{s^5s^5}{(ds^5)}^2.
 \end{aligned}
\ela
so that $M$ is foliated by self-dual Einstein spaces with $s^5$ being the foliation parameter. Moreover, the foliation by the leaves $s^5=\mbox{const}$ is dictated by the Husain-Park-type equations \eqref{F7}.  It is observed that the trichotomy encountered in the preceding and represented by the general heavenly, Husain-Park and Pleba\'nski equations is reminiscent of that observed in \cite{KonopelchenkoSchiefSzereszewski2021} in connection with the classification of eigenfunction equations associated with self-dual Einstein spaces.
\bigskip

\noindent
{\bf Acknowledgement.} B.K.\ acknowledges the financial support of the project MMNLP of the CSN IV of INFN (Italy). U.J.\ was partially supported through the International Visitor Programme of the Sydney Mathematical Research Institute (SMRI). He wishes to thank the members of the institute for their hospitality and for the inspiring research environment during his stay Jul - Sep 2023. W.K.S.\ expresses his gratitude to the SMRI for fostering the research collaboration which led to the discoveries presented in this paper. U.J.\ and W.K.S.\ also wish to thank Emma Carberry for discussions and bringing us together in the SMRI facilities. Franz Pedit's valuable insights into the subject area are also greatly appreciated.

%%%%%%%%%% Insert bibliography here %%%%%%%%%%%%%%

\end{document}